# Fermi surfaces in Kondo insulators

**Hsu Liu,**[1]* **Máté Hartstein,**[1]* **Gregory J. Wallace,**[1] **Alexander J. Davies,**[1] **Monica Ciomaga Hatnean**[2], **Michelle D. Johannes**[3], **Natalya Shitsevalova**[4], **Geetha Balakrishnan**[2] **and Suchitra E. Sebastian**[1]

[1]Cavendish Laboratory, Cambridge University, JJ Thomson Avenue, Cambridge CB3 0HE, UK,
[2]Department of Physics, University of Warwick, Coventry CV4 7AL, UK,
[3]Center for Computational Materials Science, Naval Research Laboratory, Washington, DC 20375, USA,
[4]The National Academy of Sciences of Ukraine, Kiev 03680, Ukraine.

*These authors contributed equally to this work.

E-mail: suchitra@phy.cam.ac.uk

5 March 2018

**Abstract.** We report magnetic quantum oscillations measured using torque magnetisation in the Kondo insulator $YbB_{12}$ and discuss the potential origin of the underlying Fermi surface. Observed quantum oscillations as well as complementary quantities such as a finite linear specific heat capacity in $YbB_{12}$ exhibit similarities with the Kondo insulator $SmB_6$, yet also crucial differences. Small heavy Fermi sections are observed in $YbB_{12}$ with similarities to the neighbouring heavy fermion semimetallic Fermi surface, in contrast to large light Fermi surface sections in $SmB_6$ which are more similar to the conduction electron Fermi surface. A rich spectrum of theoretical models is suggested to explain the origin across different Kondo insulating families of a bulk Fermi surface potentially from novel itinerant quasiparticles that couple to magnetic fields, yet do not couple to weak DC electric fields.

## Introduction

Evidence for a bulk Fermi surface in the Kondo insulator $SmB_6$ has been observed from a variety of experimental techniques, spanning quantum oscillations in the torque magnetisation [1, 2, 3], finite linear specific heat [1, 2, 4, 5], oscillatory magnetic entropy [2], and magnetic field enhanced thermal conductivity [2, 6]. A remarkable possibility is that the existence of a Fermi surface is more universal to correlated insulators. The observation of a Fermi surface in multiple families of correlated insulators would support a new paradigm that is distinct from the traditional idea where Fermi surfaces are the preserve of Fermi liquids. Here, we experimentally explore the possibility of a bulk Fermi surface in the Kondo insulator $YbB_{12}$, a material that is closely related to $SmB_6$ [7].

Kondo insulators are characterised by an energy gap arising from collective $f$-electron-conduction electron hybridisation, as schematically shown in figure 1(a) [8]. In the case of the Kondo insulator $YbB_{12}$, collective hybridisation occurs between the $f$-electron band and two conduction electron bands (shown in figure 4(a)), leading to a complex hierarchy of gaps. A small indirect gap of size $\approx 5$ meV in $YbB_{12}$, similar in size to that in $SmB_6$, is determined from electrical transport measurements (figures 1(a) and (b), [9, 10]), accompanied by a larger direct hybridisation gap of $\approx 200$ meV (figure 1(a)) accessed by complementary measurements such as optical spectroscopy [11] and tunneling spectroscopy [12]. Early observations of a finite linear term in the specific heat capacity (figure 1(d), [13]) and a finite density of states in tunneling experiments [10] led to broad debate about the Kondo insulating mechanism in these materials. In view of the itinerant low energy excitations revealed in the Kondo insulator $SmB_6$ by quantum oscillation measurements [1, 2, 3], we revisit this question by searching for quantum oscillations in the magnetic torque measured on single crystals of $YbB_{12}$.

## Methods

The growth of single phase $YbB_{12}$ is challenging given its peritectic phase diagram and decomposition into $YbB_{66}$ beyond a very narrow temperature range, necessitating careful control of temperature and composition during the growth process [14, 15]. Source $YbB_{12}$ powder was prepared in polycrystalline form by borothermal reduction of a mixture of $Yb_2O_3$ (99.998 mass % purity) and amorphous B (99.9 mass % purity) at 1700°C under vacuum [16]. The material was then isostatically pressed into a cylindrical rod and sintered at 1600°C in argon gas flow for several hours. Single crystal growth of $YbB_{12}$ was carried out at the University of Warwick by the traveling solvent floating zone technique under conditions similar to those previously reported in [14] using a four-mirror Xenon arc lamp (3 kW) optical image furnace (Crystal Systems Incorporated, Japan). The growths were performed in a reducing atmosphere (Ar + 3% $H_2$) at a growth rate of 18 mm/hr with the feed and seed rods counter-rotating at 20–30 rpm. $LuB_{12}$ single crystals were prepared at the National Academy of Sciences of Ukraine, Kiev by the inductive floating zone method as described in [16]. Laue x-ray imaging with a Multiwire Laue system was used to determine the quality of the grown crystal boules and to select and orient single crystal samples cut from the as-grown boule. Single crystals were selected that yielded high inverse electrical resistivity residual ratios (figure 1(b)) and well-defined spots in the Laue diffraction pattern, evidencing high single crystal quality (figure 1(c)). Elemental composition analysis was performed on selected single crystals using an FEI Philips XL30 sFEG scanning electron microscope (SEM) to reveal an atomic ratio between Yb and B closely comparable to the stoichiometric ratio of 1 : 12 and distinct from the ratios for $YbB_6$ and for $YbB_{66}$. Energy dispersive x-ray (EDX) microanalysis on multiple samples provided comparable results to SEM. Rietveld refinement performed using the Bruker TOPAS software on powder x-ray diffraction data yielded a lattice constant of 7.4686(1) Å, agreeing well with published data for $YbB_{12}$ [16].

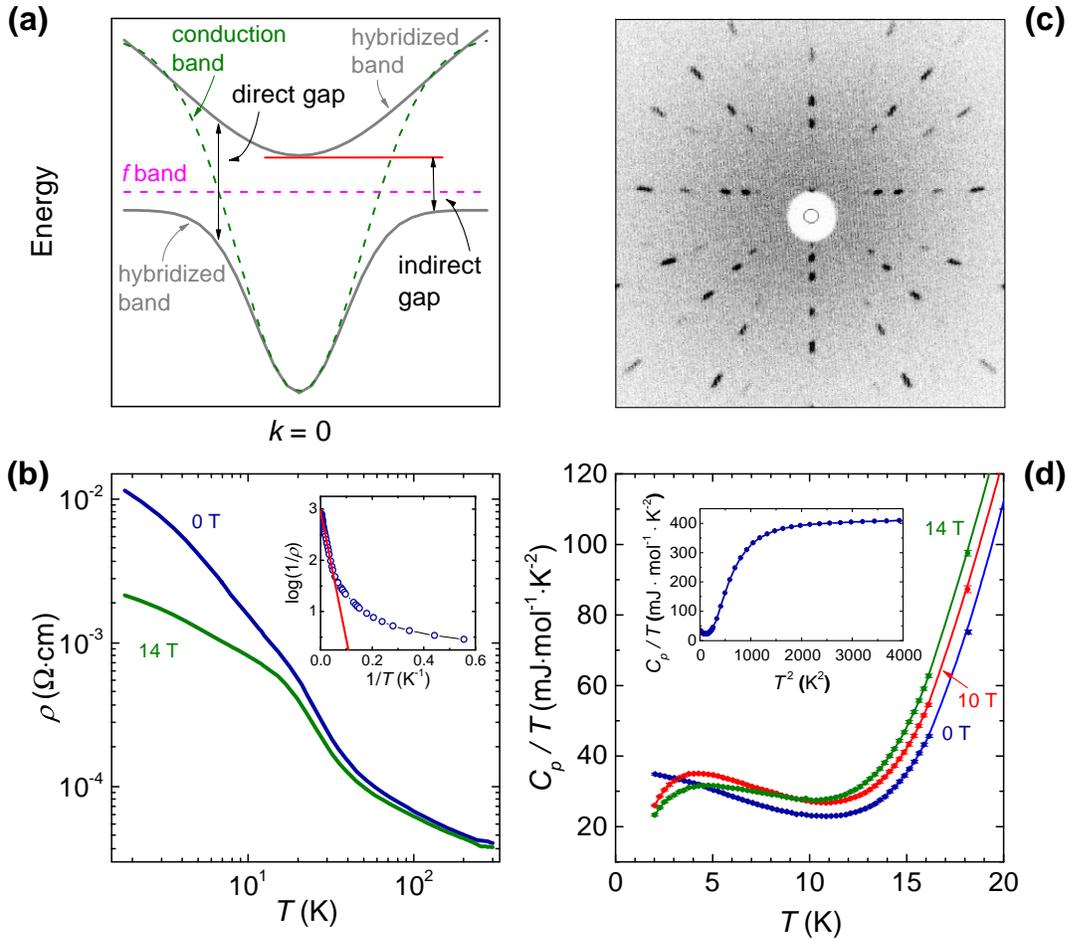

**Figure 1.** (a) Schematic depiction of the gap formation in a Kondo insulator by collective hybridisation of the $f$-electron and conduction electron bands. (b) Measured electrical resistivity as a function of temperature in magnetic fields of 0 T and 14 T. The inset shows the exponential fit $1/\rho \propto \exp\left(-2\Delta/k_BT\right)$ to the electrical resistivity $\rho$ in YbB$_{12}$ at 0 T, where $\Delta$ is the activation gap. The indirect hybridisation bandgap is found to be $\approx 5$ meV. (c) X-ray Laue back reflection photograph of a crystal of YbB$_{12}$ (on which measured quantum oscillations are shown in figure 2) along the [001] direction, showing single crystallinity. (d) Measured specific heat capacity divided by temperature of YbB$_{12}$ at 0 T, 10 T, and 14 T. A finite linear specific heat coefficient at low temperatures is seen, similar to that observed in SmB$_6$ [1, 2, 4, 5]. The measured linear specific heat coefficient is similar to that reported in [13]. While the finite value of $\gamma$ above the upturn is enhanced by a magnetic field, as would be expected for a reduction in the activation gap, the low temperature upturn in specific heat capacity is suppressed by a magnetic field, suggesting its correspondence to a secondary low energy scale such as that associated with magnetic impurities or magnetic excitons, which is suppressed by a magnetic field. The inset shows the zero-field specific heat capacity divided by temperature versus $T^2$ up to 63 K.

Torque magnetisation measurements were made on $YbB_{12}$ and $LuB_{12}$ in DC magnetic fields at the University of Cambridge (up to 14 T) and at the National High Magnetic Field Laboratory (NHMFL) in Tallahassee (up to 45 T) using flexible T-shaped BeCu cantilevers of 20 or 50 µm thickness, with the narrow end anchored and the wide end floating above a fixed Cu film. Single crystals of dimensions approximately $1 \times 1 \times 0.5$ mm$^3$ were mounted on the wide end of the cantilever. The cantilever and the Cu film form the two plates of a capacitor whose capacitance is measured using a General Radio analogue capacitance bridge in conjunction with a Stanford Research Systems lock-in amplifier. The measured change in capacitance $\Delta C$ is proportional to the change in magnetic torque. Specific heat measurements were made using the standard heat capacity option for the Physical Property Measurement System (PPMS) from Quantum Design Inc (figure 1(d)).

Density functional theory Fermi surfaces were calculated with the Wien2k augmented plane wave plus local orbital (APW+lo) code [17]. The modified Becke-Johnson (mBJ) potential was used, which is a semi-local approximation to the exact exchange plus a screening term [18] and which improves the band gap in many semiconductor materials. Application of mBJ resulted in a non-magnetic ground state with an indirect band gap of 21 meV and a direct gap of 80 meV, whereas the standard Perdew Burke Ernzerhof (PBE) potential produced a semimetal with overlapping valence and conduction bands. Spin-orbit coupling was included via the second variational method and resulted in a strong reordering of the bands. Self-consistent calculations were converged using a k-mesh of $15 \times 15 \times 15$ followed by a non-self-consistent calculation with a $30 \times 30 \times 30$ mesh for calculation of Fermi surfaces. Extremal cross-sectional areas of the Fermi surfaces were calculated for the magnetic field in the [001], [110] and [111] cubic crystal directions using the open source visualization software, OpenDX.

Effective masses were obtained by shifting the Fermi energy up/down by 0.7 meV from its original value, obtaining the new cross-sections and then calculating the cyclotron effective mass using the resulting finite differences.

## Quantum oscillations in YbB$_{12}$

Figure 2(a) shows the magnetic torque measured on a single crystal of YbB$_{12}$ as a function of DC magnetic fields up to 45 T, from which a monotonic smooth background has been subtracted. Quantum oscillations periodic in inverse magnetic field are observed for magnetic field tilt angles $\theta$ spanning [001], [111] to [110] crystal directions, and extending down to magnetic fields at least as low as 22 T. An example Fourier transform of quantum oscillations in inverse magnetic field shown in figure 2(c) reveals multiple quantum oscillation frequencies between $300 - 1500$ T. The quantum oscillation frequency $>1$ kT is seen most clearly in the high magnetic field range (figure 2(c) inset). While this higher quantum oscillation frequency is close to a harmonic of lower frequency quantum oscillations, there is no obvious observation of frequencies corresponding to harmonics of dominant amplitude low frequency quantum oscillations. The temperature dependence of the measured quantum oscillation amplitude follows a Lifshitz-Kosevich (LK) form, yielding cyclotron effective masses of $m^*/m_e$ between $3 - 10$ for the various measured quantum oscillation frequencies (figure 2(d), [19, 20]). Figure 2(b) shows the angular dependence of the measured quantum oscillation frequencies for magnetic field tilt angles spanning [001], [111] to [110] crystal directions, which reveals only a subtle variation in quantum oscillation frequency as a function of angle, consistent with a three-dimensional Fermi surface geometry.

We calibrate the absolute amplitude of quantum oscillations we observe in YbB$_{12}$ in units of $\mu_B$ per Yb unit cell by using the spring constant of the cantilever used to make torque magnetisation measurements. The method to convert the measured capacitance to absolute units of magnetic torque is the same as that detailed in [2]. Keeping the notation consistent with [2], we have cantilever length $L = 3.1$ mm, distance between cantilever and fixed Cu plate $d_0 = 0.1$ mm, spring constant $k = 190$ N $\cdot$ m$^{-1}$, lattice constant $a_{\text{u.c.}} = 0.747$ nm, and crystal volume $s^3 = 0.9 \cdot 0.5 \cdot 0.3$ mm$^3$. We thus convert the measured torque magnetisation in terms of capacitance ($C$) to an absolute magnetic moment $p_s$ in units of Bohr magneton per unit cell by the expression:

$$\Delta p_s = \frac{20}{\mu_0 \, H \sin \theta_M} \cdot \Delta C \quad \text{T} \cdot \text{pF}^{-1} \, \mu_B \text{ per unit cell.} \tag{1}$$

Here, $20$ T $\cdot$ pF$^{-1}$ is calculated with the parameters above according to [2], and $\theta_M$ is the angle between the magnetic field $\mu_0 H$ and the total magnetic moment.

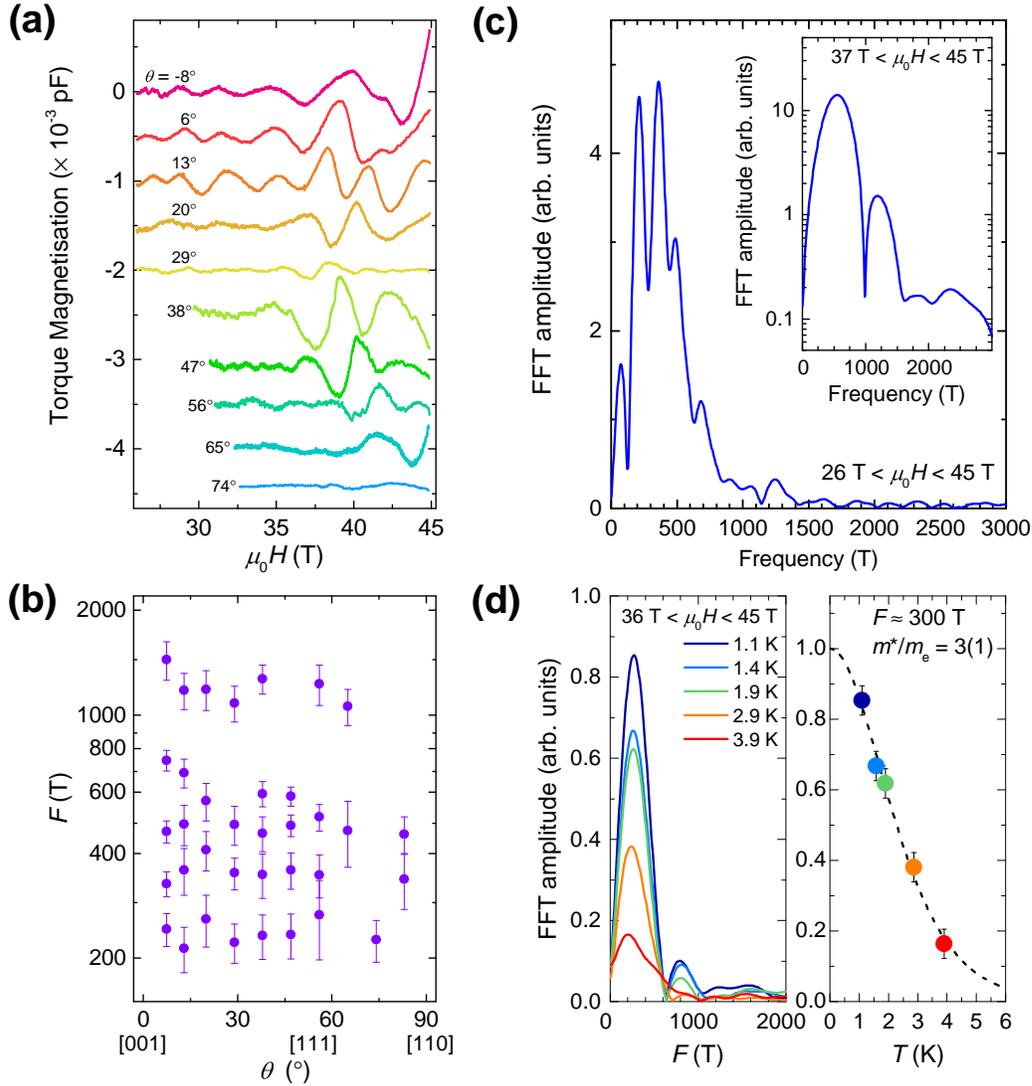

**Figure 2.** (a) Solid lines show de Haas-van Alphen oscillations measured on a single crystal of YbB$_{12}$ using torque magnetisation at $T = 0.4$ K for $\mu_0 H$ oriented at different tilt angles ($\theta$) away from the [001] crystalline direction, passing through the [111] crystalline direction, and approaching the [110] crystalline direction. (b) Angular dependence of measured quantum oscillation frequencies for values of magnetic field tilt angle spanning the [001] crystalline direction, through the [111] crystalline direction, and approaching the [110] crystalline direction. (c) Example Fourier transform of the magnetic field sweep at $\theta \approx 13°$ showing quantum oscillation peaks for a magnetic field window between 26 T and 45 T. The inset shows the Fourier transform of the magnetic field sweep at $\theta \approx 13°$ for a high magnetic field window between 37 T and 45 T, more clearly showing the quantum oscillation peak at 1.2 kT. (d) Left panel shows an example Fourier transform at $\theta \approx 6°$ measured at different temperatures. Right panel shows the quantum oscillation amplitude obtained from the peak height of the Fourier transforms shown in the left panel, plotted as a function of temperature $T$. Performing a Lifshitz-Kosevich fit (shown by a dashed line) yields an effective mass of $m^*/m_e = 3(1)$.

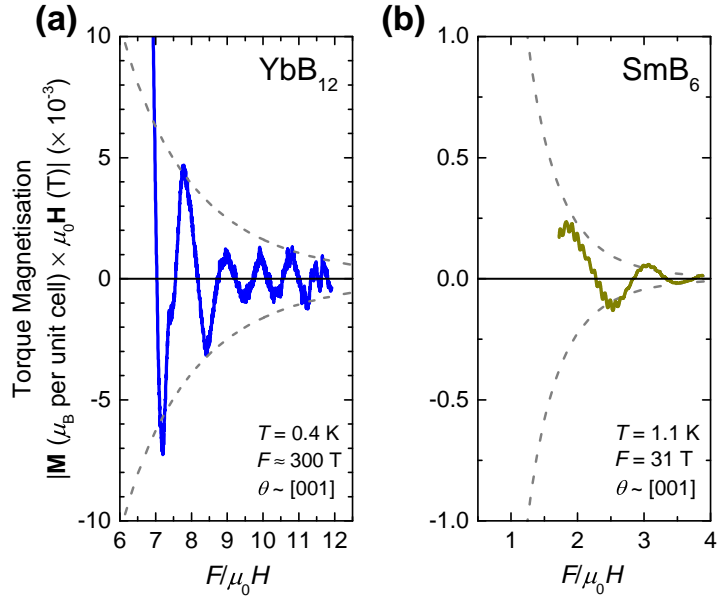

**Figure 3.** Quantum oscillatory magnetic moment (in $\mu_B$ per unit cell) corresponding to the measured oscillations versus inverse magnetic field in $YbB_{12}$ in (a), and $SmB_6$ in (b) (reproduced from [2]). Dashed lines represent the calculated magnetic field dependence of the quantum oscillation amplitude from exponential Dingle damping, for a damping factor of $\approx 100$ T in the case of $YbB_{12}$, and 30 T in the case of $SmB_6$. The theoretical Lifshitz-Kosevich estimate for the quantum oscillatory magnetic moment taking into account the angular anisotropy term, Dingle and spin-splitting damping factors is found to be $1$–$4 \times 10^{-4}$ $\mu_B$ per unit cell at $F/\mu_0 H = 6.9$ in the case of $YbB_{12}$, and $0.2$–$1 \times 10^{-4}$ $\mu_B$ per unit cell at $F/\mu_0 H = 1.9$ in the case of $SmB_6$.

We compare the measured size of the quantum oscillations to the theoretical Lifshitz-Kosevich estimate for a quantum oscillatory magnetic moment of bulk origin, including the angular anisotropy term, Dingle and spin-splitting damping factors as detailed in [2]. For the quantum oscillations observed in $YbB_{12}$, we use the experimentally measured values inferred from figure 2, and a Dingle damping factor $R_D = \exp(-100 \text{ T}/\mu_0 H)$. We find the expected theoretical amplitude of the magnetic moment for the $F \approx 300$ T frequency of $YbB_{12}$ to be of the order of $1$–$4 \times 10^{-4}$ $\mu_B$ per unit cell at $\mu_0 H = 45$ T. This theoretical estimate is comparable in order of magnitude to the experimentally measured amplitude of quantum oscillations shown in figure 3(a).

We consider a breadth of experimental observations to discern whether the observed quantum oscillations correspond to the bulk volume of the sample. Firstly, we find that the experimentally measured quantum oscillation amplitude (shown in figure 3) is in agreement with the theoretical estimate from the Lifshitz-Kosevich theory for quantum oscillations arising from the bulk volume of the sample, as calculated above. Secondly, we compare two samples with differing impurity concentrations, and examine whether the amplitude of quantum oscillations scales with the Dingle impurity term as would be expected for quantum oscillations arising from the bulk volume of the sample. We measure quantum oscillations in a single crystal of $YbB_{12}$ with more domains in the Laue pattern signaling grain boundaries or inclusions, and a lower value of inverse electrical residual resistivity ratio than the sample on which quantum oscillations are shown in figures 2 and 3. Quantum oscillations are observed on this sample, with a substantially higher Dingle impurity damping factor of $R_D = \exp(-300 \text{ T}/\mu_0 H)$. We find the amplitude of quantum oscillations observed in this single crystal with increased impurity levels to be more than an order of magnitude lower than the higher quality single crystal shown in figures 2 and 3, as expected for quantum oscillations originating from the bulk of the single crystal rather than from small secondary phase inclusions. Thirdly, we study the neighbouring semimetal $YbB_6$ in the peritectic phase diagram, to examine the likelihood of the observed quantum oscillations arising from inclusions of this phase. We find that torque magnetisation measurements on pure single crystals of $YbB_6$ yielded no discernible quantum oscillations up to an applied magnetic field of 14 T under similar experimental conditions of low temperature and high measurement sensitivity, making this an unlikely source of the quantum oscillations observed in single crystals of $YbB_{12}$. Fourthly, we note the quantum oscillation amplitude at the highest magnetic fields potentially shows a growth in amplitude beyond that expected solely from Dingle impurity damping for magnetic field tilt angles close to the [001] crystalline direction (figures 2(a) and 3(a)). Especially given that the magnetic field at which an insulator-metal transition occurs in $YbB_{12}$ is lowest ($\approx 50$ T) along the [001] crystalline direction [21], a rapid increase in quantum oscillation amplitude at high magnetic fields would be a natural consequence of the approach to an insulator-metal phase transition in bulk $YbB_{12}$, unlike in the case of a secondary phase inclusion.

**Fermi surface origin in $YbB_{12}$**

We explore band structure calculations to shed light on the origin of the Fermi surface observed in $YbB_{12}$ yielding quantum oscillations with low frequency and high effective mass, in contrast to the quantum oscillations of both low and high frequency and low effective mass observed in $SmB_6$. Band structure calculations of $YbB_{12}$ using the mBJ potential are shown in figure 4. States at both the valence band maximum and conduction band minimum are mainly derived from Yb $f$-states, but are heavily hybridised with the dispersive boron $s$ and $p$ states in the vicinity. We decompose the eigenvectors at each $k$-point both orbitally and atomically. Integrating over all $k$-points, we find, for the 14 states of the Yb $f$-complex, 13.2 below the Fermi energy and 0.8 above (figures 4(a), 4(b)). Counting the empty states indicates 0.8 holes in the $f$-complex, consistent with a nominally $Yb^{3+}$ semiconducting state that corresponds well with experiment [7] and band structure calculations that use a GW potential [22]. Interestingly, application of DFT+U, the most commonly used technique for dealing with metallic $f$-states, instead results in all 14 states being filled and corresponds to $Yb^{2+}$, which is inconsistent with experiment, as also found by previous band structure calculations [23]. A crucial distinction is therefore seen between the band structure of $SmB_6$ and $YbB_{12}$. In the case of $SmB_6$, the boron bands are filled and consequently a single half-filled unhybridised conduction $d$-electron band crosses the Fermi energy. Hybridisation of this conduction band with the $f$-electron band yields the Kondo gap. In contrast, in the case of $YbB_{12}$, two partially filled unhybridised $s$-$p$ conduction electron bands that are cumulatively half-filled cross the Fermi energy with electron-like character, and are gapped by hybridisation with the $f$-electron band as shown in figures 4(a) and 4(b).

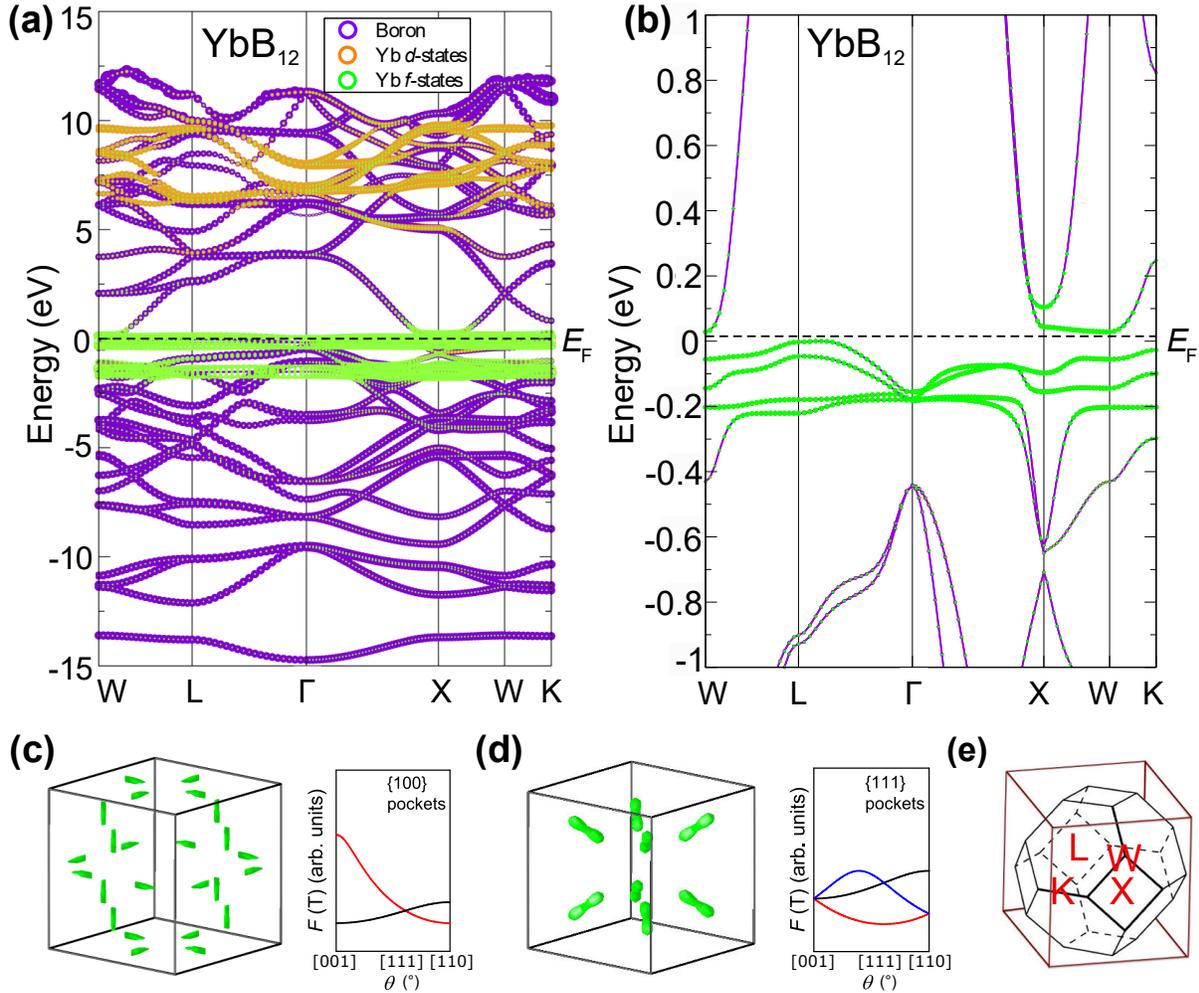

**Figure 4.** Calculated band structure of YbB$_{12}$ shown over a wide energy range in (a) with an expanded view around the Fermi energy $E_F$ in (b). Several characters are projected out of the eigenvectors at each $k$-point and the resulting weight is indicated by a circle of proportional size. Green circles are Yb $f$-states, orange circles are Yb $d$-states, and violet circles are boron states. (c) Small needle-shaped Fermi surfaces of YbB$_{12}$ obtained using the modified Becke-Johnson potential for a small positive energy shift. Expected angular dependence of the quantum oscillation frequencies in the [001]-[111]-[110] rotation plane are shown by approximating the shown Fermi surfaces as prolate ellipsoids. (d) Small peanut-shaped Fermi surfaces of YbB$_{12}$ obtained using the modified Becke-Johnson potential for a small negative energy shift. Expected angular dependence of the quantum oscillation frequencies in the [001]-[111]-[110] rotation plane are shown by approximating the shown Fermi surfaces as prolate ellipsoids. (e) A schematic of the conventional face-centred cubic Brillouin zone used for the band structures within the cubic Brillouin zone used for the Fermi surfaces.

Given the absence of a finite electronic density of states at the Fermi energy, constituent neutral quasiparticles have been proposed to explain the observation of a bulk Fermi surface in Kondo insulators. Neutral quasiparticles invoked by various theoretical models include spinons in the case of single band Mott insulating organic spin liquids [24, 25, 26, 27], and in the case of single band Kondo insulators such as $SmB_6$, magnetic excitons [28], composite excitons [29, 30], Majorana fermions [31, 32, 33], and others [34, 35]. A natural way to think of a Fermi surface of such neutral quasiparticles is in terms of slow fluctuations in space and time between the insulating ground state where a Fermi surface is absent due to filling of the Brillouin zone, and the neighbouring metallised ground state in phase space, which is characterised by a Fermi surface [1, 2]. The character of such a neutral Fermi surface may thus be expected to be akin to the Fermi surface of the neighbouring metallised ground state. Metallisation in the case of $SmB_6$ requires a decoupling of the $f$-electron and conduction electron bands, resulting in a solely conduction electron Fermi surface occupying half the Brillouin zone. Accordingly, comparison of the observed Fermi surface in insulating $SmB_6$ with a conduction electron Fermi surface similar to that in metallic $LaB_6$ was found to yield good agreement both in frequency and effective mass [1, 2]. In contrast, metallisation of $YbB_{12}$ can arise through two routes: either (i) a decoupling of the $f$-electron and conduction electron bands yielding a Fermi surface corresponding to the conduction electron band, or (ii) an effective relative shift of each of the two hybridised conduction electron bands, yielding a Fermi surface corresponding to a heavy fermion semimetal [29, 30].

We consider the first scenario, and make a comparison of the small heavy Fermi surface sections observed in insulating $YbB_{12}$ with a conduction electron Fermi surface similar to that in metallic $LuB_{12}$. In $LuB_{12}$, all 14 $f$-states are well below the Fermi energy in the calculated band structure, resulting in a metallic ground state with highly dispersive boron bands and light carriers (figure 5(a), [36, 37]). Quantum oscillations we measure using torque magnetisation in $LuB_{12}$ (shown in figures 5(e)–(g)) yield good agreement with the calculated band structure (figures 5(a)–(d)), and with previous quantum oscillation measurements [36, 38]. A correspondence is not immediately obvious between the observed quantum oscillations with a relatively high effective mass in $YbB_{12}$ (figure 2(d)) and the large and light conduction electron Fermi surface expected from band structure and observed in $LuB_{12}$ (figure 5). It is possible that owing to subtle materials differences between $YbB_{12}$ and $LuB_{12}$, a band shift could yield small Fermi surface pockets from the conduction electron band similar to those observed in $YbB_{12}$. Meanwhile quantum oscillations corresponding to large Fermi surface pockets might not be observed due to even higher effective masses than for the small Fermi surfaces observed. Alternatively, in the second scenario, metallisation in $YbB_{12}$ can be achieved by a small effective relative shift of the two conduction bands while retaining hybridisation or potentially reduced hybridisation. In this case, small heavy Fermi surface sections characteristic of a heavy fermion semimetal [8] would be expected, as can be seen from performing small energy shifts to the band structure (figures 4(c) and (d)). Such heavy Fermi surface sections would also yield a sizable linear heat capacity as observed in $YbB_{12}$ (figure 1(d), [13]). This scenario of a neutral Fermi surface in Kondo insulating $YbB_{12}$ with properties similar to the Fermi surface of a heavy fermion semimetal, arising from a small relative shift to the two hybridised conduction bands, is intriguing to pursue theoretically. The realm of Fermi surfaces in Kondo insulators may be even richer than previously thought. Just as materials differences between metals yield differences in band structure and consequently Fermi surface character, we find that differences between the nature of $f$-electron and conduction electron hybridisation in $SmB_6$ and $YbB_{12}$ yield potentially important differences in Fermi surface character between the two Kondo insulators.

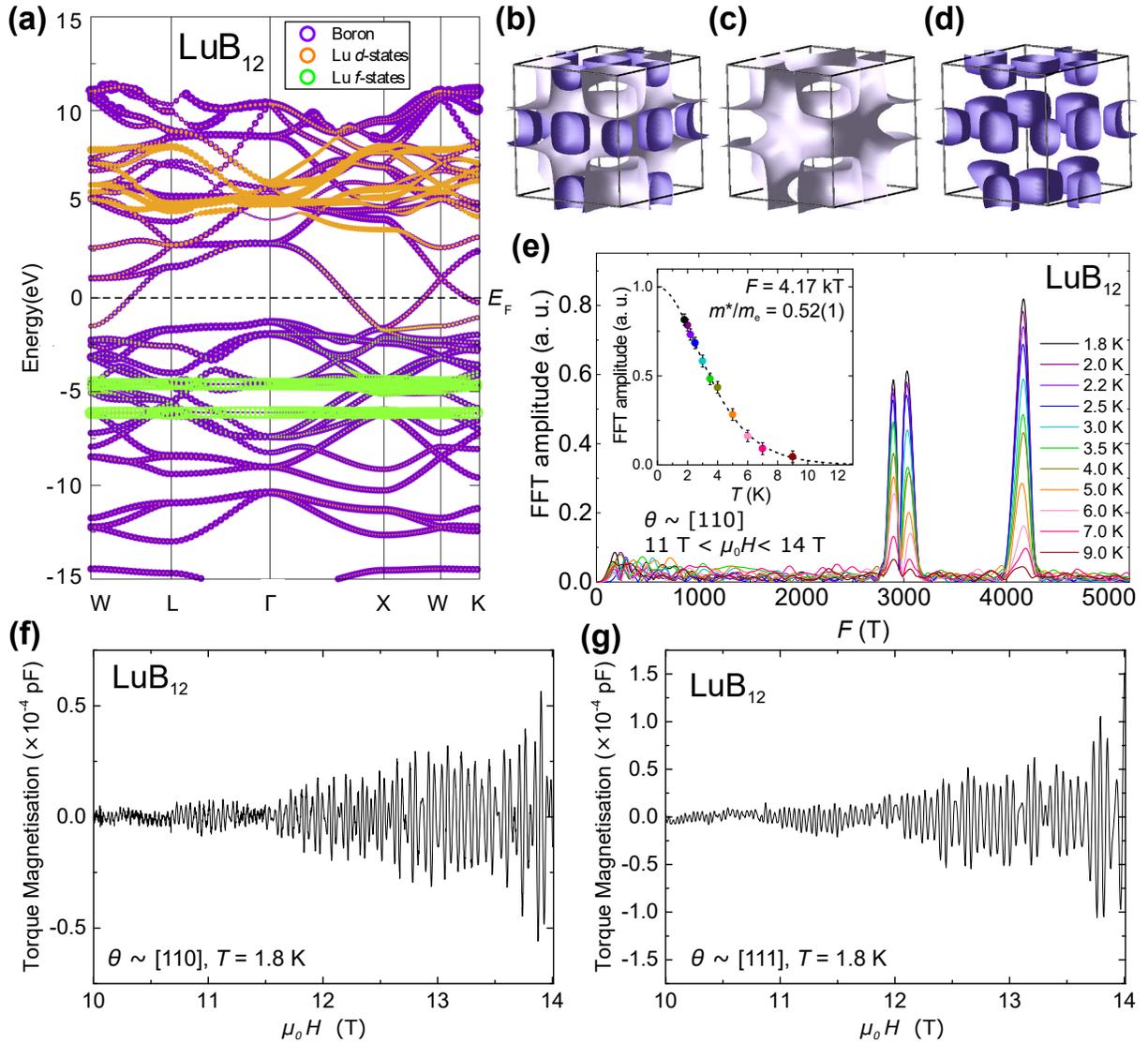

**Figure 5.** (a) Calculated band structure of LuB$_{12}$. Several characters are projected out of the eigenvectors at each k-point and the resulting weight is indicated by a circle of proportional size. Green circles are Lu $f$-states, orange circles are Lu $d$-states, and violet circles are boron states. (b) - (d) the Fermi surfaces of LuB$_{12}$ shown together in (b) and separately for clearer viewing in (c) and (d). (f) - (g) De Haas-van Alphen oscillations measured using torque magnetisation at 1.8 K in LuB$_{12}$. (e) Fourier transform of the magnetic field sweeps taken at different temperatures. The inset shows the Lifshitz-Kosevich fit (dashed line) to the frequency $F = 4.17$ kT. The effective masses $m^*/m_e$ are found to be 0.40(1), 0.53(2), and 0.52(1) for the frequencies 2.90 kT, 3.03 kT, and 4.17 kT for $\mu_0 H$ oriented a few degrees away from [110], corresponding well to calculated frequencies 2.92 kT and 3.93 kT for $\mu_0 H \parallel$ [110] with effective masses $m^*/m_e$ of 0.36 and 0.43 respectively. Frequencies 2.89 kT, 3.51 kT, 3.75 kT, and 5.64 kT are measured for $\mu_0 H$ oriented a few degrees away from [111], comparable to calculated frequencies of 2.82 kT and 5.80 kT for $\mu_0 H \parallel$ [111] with effective masses $m^*/m_e$ of 0.35 and 0.67 respectively.

**Magnetic field tuning in $YbB_{12}$**

Another salient difference between $SmB_6$ and $YbB_{12}$ is the effectiveness of magnetic field in tuning these materials toward an insulator-metal transition. While applied magnetic fields as high as $\mu_0 H = 93$ T only result in a negative magnetoresistance of $\approx 7$ % in $SmB_6$ [39], applied magnetic fields of $\mu_0 H \approx 45$ T are found to reduce the electrical resistivity by an order of magnitude in $YbB_{12}$ (figure 1(b), [21]). The enhanced response of $YbB_{12}$ to applied magnetic fields compared to $SmB_6$ may reflect small band shifts due to Zeeman splitting or the involvement of magnetic degrees of freedom in the ground state of $YbB_{12}$. The close proximity of $YbB_{12}$ to the insulator-metal transition may be expected to lead to an increased propensity for quantum oscillations originating from neutral quasiparticles [24, 25, 26, 27, 29, 30, 31, 32, 33]. Such an effect would yield larger amplitude quantum oscillations at higher magnetic fields, potentially yielding the increase in quantum oscillation amplitude beyond that expected from Dingle damping suggested from the experimental data for magnetic field tilt angles near the [001] crystalline direction in $YbB_{12}$ (figures 2(a) and 3(a)). Alternatively, the reduced charge gap at high magnetic fields could also provide an explanation in terms of conventional quasiparticles tunneling through a narrow energy gap that are more likely to yield quantum oscillations [40, 41, 42, 43]. Taken in conjunction with the observation of a finite linear specific heat capacity (figure 1(d), [13]) even in zero magnetic field, an explanation for the observed quantum oscillations in terms of a Fermi surface originating from novel quasiparticles that couple to a magnetic field but not to a weak DC electric field appears more likely than an origin from conventional tunneling through a magnetic field reduced charge gap.

## Summary

The observation of magnetic quantum oscillations in at least two families of Kondo insulators, $SmB_6$ and $YbB_{12}$, suggests a more universal phenomenon across correlated insulators. Differences we uncover between the character of the underlying bulk Fermi surface in the two systems further add to the richness of potential Fermi surface models relevant to the broad panorama of Kondo insulators. While the Fermi surface observed in $SmB_6$ corresponds in geometry and effective mass to the conduction electron Fermi surface, the Fermi surface observed in $YbB_{12}$ corresponds more closely to that of a heavy fermion semimetal, suggesting important differences between theoretical models of relevance to each of these systems. The magnetic field tuning we find to influence the quantum oscillations observed in $YbB_{12}$, while having little effect on the quantum oscillations in $SmB_6$, further informs our understanding of the itinerant low energy excitations involved in each of these materials and their approach to the neighbouring insulator-metal quantum critical point. Looking ahead, magnetic field dependent quantum oscillation measurements in other quantities, as well as complementary measurements such as thermal conductivity and nuclear magnetic resonance at low temperatures, are important in the quest to examine the character of the novel itinerant low energy excitations in $YbB_{12}$. Our results provide further impetus to the search underway for new theoretical paradigms to explain the unexpected discovery of a bulk Fermi surface in Kondo insulating materials [24, 25, 26, 27, 29, 30, 31, 32, 33, 40, 41, 42, 43].

## Acknowledgments

We thank the team at the National Academy of Sciences of Ukraine, Kiev for assistance in the preparation of polycrystalline $YbB_{12}$ as well as single crystals of $LuB_{12}$. We are grateful to Y. Matsuda for discussions of their unpublished measurements on $YbB_{12}$. We acknowledge valuable discussions with G. G. Lonzarich, T. Senthil, D. Chowdhury, I. Sodemann, and G. Baskaran. We are grateful for the experimental support provided by the NHMFL, Tallahassee, including J. Billings, J. T. Camacho, R. Carrier, E. S. Choi, W. A. Coniglio, B. L. Dalton, D. Freeman, L. J. Gordon, M. Hicks, S. A. Maier, T. P. Murphy, J.-H. Park, J. N. Piotrowski, J. A. Powell, E. Stiers.

H.L, M.H., G.W., A.J.D, and S.E.S. acknowledge support from the Royal Society, the Leverhulme Trust through the award of a Philip Leverhulme Prize, the Winton Programme for the Physics of Sustainability, EPSRC UK (grant number EP/M000524/1) and the European Research Council under the European Unions Seventh Framework Programme (grant number FP/2007-2013)/ERC Grant Agreement number 337425. M.C.H. and G.B. acknowledge financial support from the EPSRC UK (grant number EP/M028771/1). M.D.J. acknowledges support for this project by the Office of Naval Research (ONR) through the Naval Research Laboratory's Basic Research Program. A portion of this work was performed at the National High Magnetic Field Laboratory, which is supported by National Science Foundation Cooperative Agreement No. DMR-1157490, the State of Florida, and the Department of Energy (DOE).